\documentclass[]{aastex631}

\usepackage[textsize=scriptsize]{todonotes}
\usepackage[version=4]{mhchem}
\usepackage{physics}
\usepackage{comment}
\usepackage{amsmath}

\newcommand{\nuebar}{\bar{\nu}_e}
\newcommand{\nue}{\nu_e}
\newcommand{\nux}{\nu_x}
\newcommand{\nuxbar}{\bar{\nu}_x}
\newcommand{\numu}{\nu_\mu}
\newcommand{\nutau}{\nu_\tau}
\newcommand{\Msun}{M_{\odot}}
\newcommand{\kpc}{\mathrm{kpc}}
\newcommand{\erg}{\mathrm{erg}}
\newcommand{\MeV}{\mathrm{MeV}}

\newcommand{\yr}{\mathrm{years}}

\newcommand{\cm}{\mathrm{cm}}
\newcommand{\CPP}{C\nolinebreak\hspace{-.05em}\raisebox{.4ex}{\tiny\bf +}\nolinebreak\hspace{-.10em}\raisebox{.4ex}{\tiny\bf +}}

\newcommand{\GF}{G_{\mathrm{F}}}

\shorttitle{Supernova simulator: SKSNSim}
\shortauthors{Nakanishi et al.}
\graphicspath{{./}{figures/}}

\begin{document}

\title{Supernova burst and Diffuse Supernova Neutrino Background simulator\\for Water Cherenkov Detectors\footnote{Released on March, 1st, 2021}}

\correspondingauthor{Fumi Nakanishi}
\email{nakanishi-suv@s.okayama-u.ac.jp}

\author[0000-0003-4408-6929]{Fumi Nakanishi}
\affiliation{Department of Physics, Okayama University \\
3-1-1 Tsushima-naka, Kita-ku, Okayama\\
Okayama 700-8530, Japan}
\author[0000-0002-0808-8022]{Shota Izumiyama}
\affiliation{Depertment of Physics, Tokyo Institute of Technology,\\
2-12-1 H-26, Oookayama, Meguro-ku,\\
Tokyo, 152-8551, Japan}
\author[0000-0003-3273-946X]{Masayuki Harada}
\affiliation{Department of Physics, Okayama University \\
3-1-1 Tsushima-naka, Kita-ku, Okayama\\
Okayama 700-8530, Japan}
\author[0000-0003-0437-8505]{Yusuke Koshio}
\affiliation{Department of Physics, Okayama University \\
3-1-1 Tsushima-naka, Kita-ku, Okayama\\
Okayama 700-8530, Japan}
\affiliation{Kavli Institute for the Physics and Mathematics of the Universe (WPI),\\
The University of Tokyo Institutes for Advanced Study, University of Tokyo,\\
Kashiwa, Chiba 277-8583, Japan}



\begin{abstract}

    If a galactic supernova explosion occurs in the future, it will be critical to rapidly alert the community to the direction of the supernova by utilizing neutrino signals in order to enable the initiation of follow-up optical observations. In addition, there is anticipation that observation of the diffuse supernova neutrino background will yield discoveries in the near future, given that experimental upper limits are approaching theoretical predictions.
    We have developed a new supernova event simulator for water Cherenkov neutrino detectors, such as the highly sensitive Super-Kamiokande. This simulator calculates the neutrino interaction in water for the two types of supernova neutrinos described above. Its purpose is to evaluate the precision in determining the location of supernovae and to estimate the expected number of events related to the diffuse supernova neutrino background in Super-Kamiokande.
    In this paper, we describe the basic structure of the simulator and its demonstration.

\end{abstract}

\keywords{Neutrino telescopes (1105) -- Supernova neutrinos (1666) -- Astronomy software (1855)}

\section{Introduction}\label{sec:intro}

A core-collapse supernova (CCSN) is an explosion at the end of the evolution of a massive star and one of the most energetic astrophysical phenomena.
It releases energy of $\sim 10^{53}~\erg$ in total, $99\%$ of which is released as neutrinos.
A supernova, SN1987A, is the latest event that has been seen by eyes on the Earth.
At that time, neutrino detectors, the Kamiokande-II~\citep{1987PhRvL..58.1490H}, the IMB~\citep{1987PhRvL..58.1494B}, and the Baksan experiments~\citep{1988PhLB..205..209A}, observed neutrinos.
Since these are the first and last so far observations of neutrinos from supernova, the mechanism of the explosion is not yet well understood.
Due to the limitation of neutrino observations, studies to reveal the mechanism are based on state-of-the-art numerical simulations.
Early studies assumed the spherical symmetry of stars, for example, in the Wilson model~\citep{1998ApJ...496..216T}.
Thanks to the recent development in computers, multi-dimensional simulations have been developed and are available~\citep{2012ApJ...749...98T, 2013ApJ...770...66H, 2014PhRvD..90d5032T}.
While simulation studies have made great progress, there are growing expectations for the next observations of CCSN neutrinos. 
There are several large-volume neutrino telescopes in the world, such as Super-Kamiokande (SK)~\citep{2003NIMPA.501..418F}, IceCube~\citep{2017JInst..12P3012A}, KM3NeT~\citep{2016JPhG...43h4001A}, and KamLAND~\citep{2014EPJC...74.3094S}. There are also alert systems for follow-up observatories using these telescopes.
For example, the SK has a real-time burst monitor and an alert system~(\cite{2016APh....81...39A}; Y.Kashiwagi et al. 2024, in preparation) 
via the Global Coordinate Network~\citep{GCNWebsite}.
In addition, those large-volume neutrino detectors are coordinated in the SuperNova Early Warning System (SNEWS)~\citep{2021NJPh...23c1201A}.
Furthermore, unlike burst signals, it is thought that neutrinos generated by past supernovae have occurred in large numbers throughout the history of the universe. 
The flux from these is called Diffuse Supernova Neutrino Background (DSNB) or Supernova Relic Neutrinos (SRN)~\citep{1997APh.....7..125M, 1997APh.....7..137H}.
To further improve the sensitivity to the DSNB observation, SK was upgraded by adding gadolinium (\ce{Gd}), which has the largest thermal neutron capture cross section among all elements, to the water in the detector. This upgrade is called SK-Gd~\citep{2022NIMPA102766248A}.

We have developed a Super-Kamiokande SuperNova Simulator (SKSNSim)\footnote{Available on the GitHub, \url{https://github.com/SKSNSim/SKSNSim}.\label{fot:githab}} to evaluate the detector response and sensitivity for supernova explosions (SN burst) and DSNB.\@
SKSNSim, originally developed for the SK, is a simulator of neutrino interactions in water Cherenkov neutrino detectors and applies to any water medium detectors.
In this paper, we describe the details of this simulator.
Section~\ref{sec:structure} shows the basic simulation strategy and flow.
We show the demonstration of simulation and its usage in Section~\ref{sec:demo}.
Finally, we summarize in Section~\ref{sec:summary}.

\section{Simulation Structures}\label{sec:structure}

SKSNSim is a simulator that calculates and outputs the kinematics of particles generated by neutrino interactions from the input flux of SN burst or DSNB\@.
In this section, we explain the structure and flow of the simulation, consisting of neutrino fluxes, neutrino oscillation effects on propagation from the center of the star to the surface, and cross sections of neutrino with water.
Because pipelines for the SN burst and the DSNB share some parts, like cross sections and methods to generate particles, SKSNSim is constructed for both cases.

We have implemented the main parts of SKSNSim with \CPP\ programming language.
The cross section model and output formats are modularized, making it easy to use them in external programs and update the models.


\subsection{Simulation Flow}\label{subsec:flow}

The software has two modes for the SN burst and the DSNB simulation.
While some parts are common for both modes, there is a major difference: the SN burst flux is a function of time with sub-second units, whereas the DSNB flux is constant in time.

The SN burst simulation consists of the following steps;
(1) calculating the average number of interacting neutrinos for each neutrino type in the defined time and energy bins from the neutrino flux, cross section, and neutrino oscillation effects,
(2) throwing a random generator to get the number of interacting neutrinos according to a Poisson distribution,
(3) determining the kinematics of particles generated by the neutrino interaction, such as vertex and momentum, generated in each event using random numbers,
and (4) repeating from step 1 over the burst duration.

The following steps are used for the DSNB simulation:
(1) throwing a random generator to get the neutrino's energy,
(2) determining the kinematics of particles generated in each event using random numbers,
(3) repeating from step 1 for the number specified by users. 
The number of necessary events often generates far more than DSNB prediction. In this case, the final number of events must be normalized by the observation time.
In contrast to the SN burst steps, the DSNB steps do not include the neutrino oscillation effect in SKSNSim because most DSNB models include it already.

\subsection{Input of Supernova Model}\label{subsec:snmodel}

There are a variety of models, both for the SN burst and the DSNB, based on different simulation approaches. In dealing with any model, SKSNSim requires the original neutrino flux information for each time and energy in the SN burst simulation. In the case of DSNB simulation, the neutrino flux information for each energy is necessary. The details about the input model in the SN burst, and the DSNB simulation are written in the following sections.

\subsubsection{SN burst}\label{section:subsub_inputformat_snburst}

Many SN burst modelers publish their results with their data format to express the simulation results efficiently.
We decided to support only the text format distributed by \cite{2013ApJS..205....2N} in the SN burst simulation to handle all models in a unified manner. This format includes differential neutrino number flux and differential neutrino luminosity for every time and energy bin and types of neutrino $\nue, \nuebar, \nux, \nuxbar$.
Here, $\nux$ represents $\numu$ ($\nutau$) because the distributions of $\numu$ and $\nutau$ are equal, and also $\numu$ and $\nutau$ of $\order{10}~\MeV$ behave the same way in water Cherenkov detectors.
The conversion to the input data format is detailed in~Y.Kashiwagi et al.\ (2024, in preparation). 
SN burst models listed in Table~\ref{tab:snburst_model} are implemented in SKSNSim.
\begin{deluxetable*}{lll}
    \tablecaption{List of current supported SN burst models. \label{tab:snburst_model}}
    \tablehead{\colhead{Model} & \colhead{Supported configuration} & \colhead{Reference}}
    \startdata
    Nakazato & all parameters in the reference & \citet{2013ApJS..205....2N} \\
    Mori & $9.6~\Msun$ & \citet{2021PTEP.2021b3E01M} \\
    Wilson & $20~\Msun$ & \citet{1998ApJ...496..216T} \\
    Tamborra & $27~\Msun$ and ``black'' observer direction & \citet{2014PhRvD..90d5032T} \\
    Fischer & $8.8~\Msun$ & \citet{2010AaA...517A..80F} \\
    H\"{u}depohl & full neutrino interactions and $8.8~\Msun$ & \citet{1987ApJ...322..206N} \\
    \enddata
    \tablecomments{Except for the Nakazato and Mori models, the flux variation is converted to the format that \cite{2013ApJS..205....2N} defined.}
\end{deluxetable*}

\subsubsection{DSNB}\label{section:subsub_inputformat_dsnb}

The time dependence of the DSNB flux is negligible for a typical observation period of $\sim 10~\yr$, such as for SK\@.
Therefore, SKSNSim simulates DSNB events using only an energy spectrum written in text format without any time dependence.
In addition, the DSNB simulation processes only $\nuebar$ flux since a dominant interaction channel in water is inverse beta decay (IBD) in the typical energy region of the DSNB at a few tens of MeV\@.
Users can specify any binned $\nuebar$ spectrum to simulate the IBD interaction via a text file.
With this advantage, the flux of other $\nuebar$ sources, such as nuclear reactors, can be simulated with this tool. By default, it supports the $\nuebar$ flux model provided by \cite{2009PhRvD..79h3013H}. 
Another characteristic of SKSNSim is that it is implemented to simulate events according to a flat energy spectrum in a positron energy space as well as a neutrino energy space.
This implementation is useful in a ``re-weighting'' method.
We explain in detail the use of the re-weighting method for the SK DSNB analysis in Section~\ref{subsec:use_sk}.

\subsection{Neutrino Oscillation}\label{subsec:nuosc}
In SKSNSim, we consider only the Mikheyev-Smirnov-Wolfenstein (MSW) effect~\citep{1978PhRvD..17.2369W, 1985YaFiz..42.1441M} as the effect of neutrino oscillation. There are also collective oscillations~\citep{2010ARNPS..60..569D} and an oscillation during propagation in a vacuum.
However, these two oscillations are not taken into consideration during an SN burst because no method can easily implement the collective oscillation, and the vacuum oscillation is negligible compared to the MSW.
As described above, the neutrino oscillation effect is not included in the DSNB simulation.
This section describes the treatment and implementation of the MSW effect in the simulation of an SN burst.

The following equations represent the neutrino flux, including the MSW effect on the surface of the star ($N_{\bullet}^{\rm{sur}}$)~\citep{2000PhRvD..62c3007D}.
Here, $N_{\nue}^{\text{gen}},~N_{\nuebar}^{\text{gen}},~\text{and}~N_{\nux}^{\text{gen}}$ represent the neutrino flux of $\nue$, $\nuebar$, and $\nux$ at generation, respectively. In this notation, $N_{\nux}^{\rm{gen}}$ expresses the flux of $\numu$ or $\nutau$. The neutrino flux on the surface of the star in the normal mass ordering case shows
\begin{equation}
    \label{equ:normal}
    \begin{split}
        N_{\nue}^{\rm{sur}} &= N_{\nux}^{\rm{gen}}\\
	    N_{\numu}^{\rm{sur}} + N_{\nutau}^{\rm{sur}} &= N_{\nue}^{\rm{gen}} + N_{\nux}^{\rm{gen}}\\
	    N_{\nuebar}^{\rm{sur}} &= N_{\nuebar}^{\rm{gen}} \times \cos^2 \theta_{12} + N_{\nuxbar}^{\rm{gen}} \times \sin^2 \theta_{12}\\
        N_{\bar{\nu}_{\mu}}^{\rm{sur}} + N_{\bar{\nu}_{\tau}}^{\rm{sur}} &= N_{\nuebar}^{\rm{gen}} \times \sin^2 \theta_{12} + N_{\nuxbar}^{\rm{gen}} \times (1+ \cos^2 \theta_{12}).
    \end{split}
\end{equation}
In contrast, in the inverted mass ordering case, the neutrino flux represents
\begin{equation}
    \label{equ:inverted}
    \begin{split}
        N_{\nue}^{\rm{sur}}  &= N_{\nue}^{\rm{gen}} \times \sin^2 \theta_{12} + N_{\nux}^{\rm{gen}} \times \cos^2 \theta_{12}\\
	    N_{\numu}^{\rm{sur}} + N_{\nutau}^{\rm{sur}} &= N_{\nue}^{\rm{gen}} \times \cos^2 \theta_{12} + N_{\nux}^{\rm{gen}} \times (1+\sin^2 \theta_{12})\\
	    N_{\nuebar}^{\rm{sur}} &= N_{\nuxbar}^{\rm{gen}}\\
	    N_{\bar{\nu}_{\mu}}^{\rm{sur}} + N_{\bar{\nu}_{\tau}}^{\rm{sur}} &= N_{\nuebar}^{\rm{gen}} + N_{\nuxbar}^{\rm{gen}}.
    \end{split}
\end{equation}

In SKSNSim, the oscillation effect is considered by multiplying the expected value by the above formula. The oscillation effect (normal mass ordering, inverted mass ordering, and no oscillation) can be selected by the user.

\subsection{Neutrino Interaction}\label{subsec:nuint}

\begin{figure}
    \centering
    \includegraphics[width=6cm]{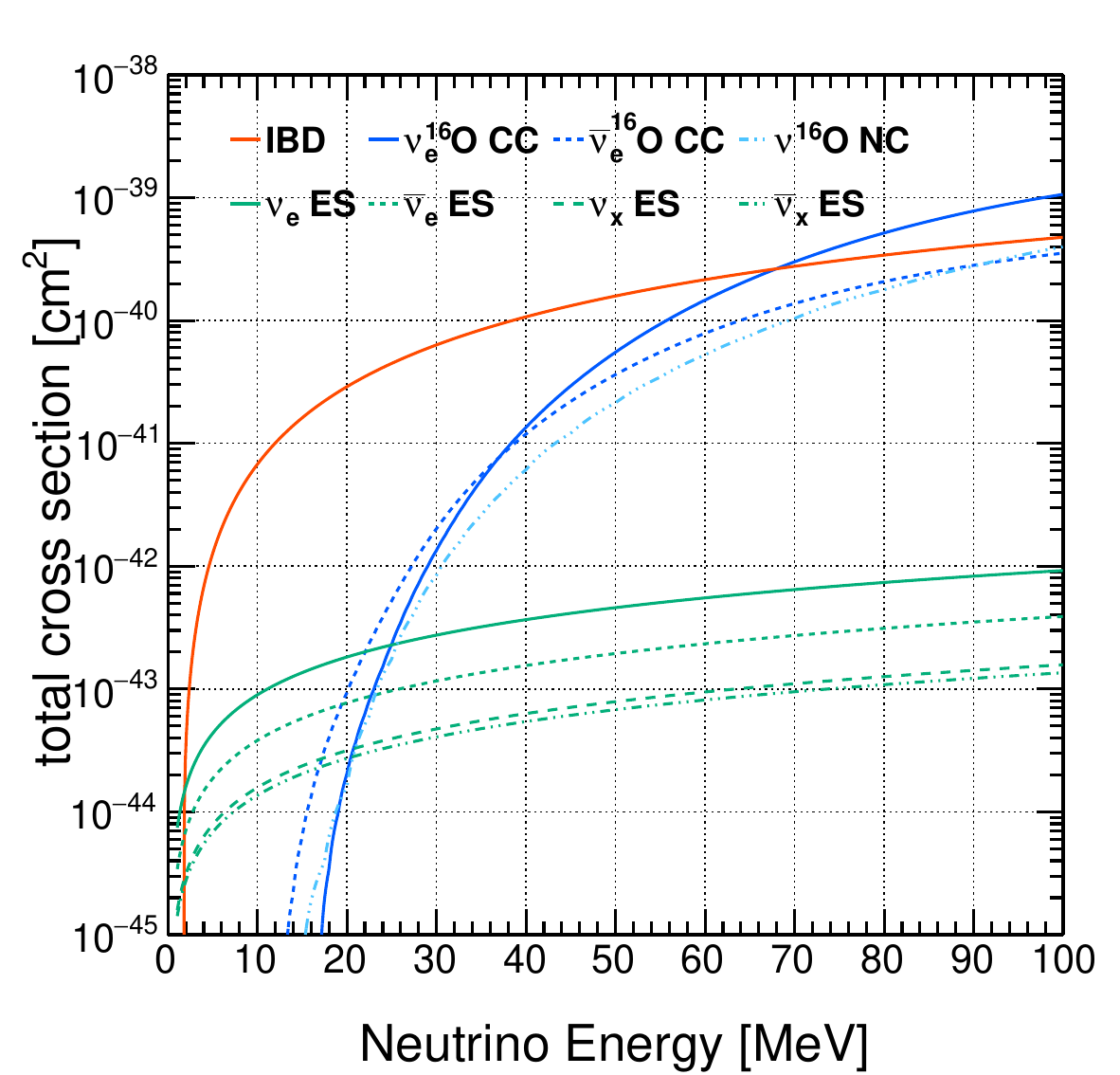}
    \caption{Total cross sections of neutrinos with water as a function of neutrino energy. The solid red indicates IBD, and the green lines represent ES with a neutrino flavor: $\nue$ (solid), $\nuebar$ (dotted), $\nux$ (dashed), and $\nuxbar$ (dot-dashed). The solid blue and dashed blue lines represent \ce{^16O} CC $\nue$ and $\nuebar$ respectively. The dot-dashed light blue line indicates \ce{^16O} NC interaction. The cross sections are calculated according to \cite{2003PhLB..564...42S} for IBD, \cite{1995PhRvD..51.6146B} for ES, \cite{2018PTEP.2018l3E02N} for \ce{^16O} CC interaction, and \cite{1996PhRvL..76.2629L} and \cite{2002PhRvD..66a3007K} for \ce{^16O} NC interaction.}\label{fig:osc_template}
\end{figure}

We consider the following four types of neutrino interactions in SKSNSim with a large cross section in the supernova neutrino energy region observed in water.
Figure~\ref{fig:osc_template} shows the cross section for each interaction.
\begin{subequations}
    \begin{align}
        \nuebar+p &\rightarrow n+e^+ &&\text{Inverse Beta Decay},\\
        \nue/\nuebar/\nu_x/\nuxbar +e^-&\rightarrow \nue/\nuebar/\nu_x/\nuxbar+e^- &&\text{Electron scattering},\\ 
        \nue/\nuebar+\mathrm{^{16}O}&\rightarrow e^-/e^++\mathrm{^{16}F/^{16}N} &&\text{Charged-current reaction},\\
    \nue/\nuebar/\nu_x/\nuxbar+\mathrm{^{16}O}&\rightarrow p/n + \gamma + \mathrm{^{15}N/^{15}O} &&\text{Neutral-current reaction}.
    \end{align}
\end{subequations}

Inverse beta decay (IBD) reaction has the largest cross section for the neutrinos with water.
The cross section of IBD is calculated as below based on \cite{2003PhLB..564...42S},
\begin{equation}
    \dv{\sigma}{t}=\frac{{\GF}^2\cos^2\theta_{\mathrm{C}}}{2\pi{(s-m_p^2)}^{2}}|\mathcal{M}^{2}|,
\end{equation}
where $\GF$ is the Fermi coupling constant, $\theta_{\mathrm{C}}$ is the Cabibbo angle, $s$ and $t$ are the Mandelstam variables which are functions of each particle's momentum, $m_p$ is the proton mass, and $|\mathcal{M}|$ is a matrix element, respectively. The order of the IBD cross section is $10^{-41}~\cm^2$ for the typical supernova neutrino energies.
SKSNSim includes the calculation by~\cite{1999PhRvD..60e3003V}, ~\cite{2003PhLB..564...42S} (default in SKSNSim), and \cite{2022JHEP...08..212R}.
Users can switch between these calculations.

Electron scattering (ES) is a reaction in which a neutrino scatters with an electron, all flavors of neutrino contribute to this interaction.
The order of the cross section is two orders of magnitude lower than the IBD cross section. However, IBD has almost no directional sensitivity to the supernova, whereas ES is an important reaction for determining the direction of the supernova in terms of having a strong directional correlation between neutrinos and scattered electrons. 
The basic equation for the reaction cross section is derived from the weak interaction, and this generator uses the following equation that takes radiative corrections into account according to \cite{1995PhRvD..51.6146B},
\begin{equation}\label{equ:crs_es}
        \dv{\sigma}{T} = \frac{2{\GF}^2 m_e}{\pi}  \left\{ {g_L}^2(T)\qty[1+\frac{\alpha}{\pi}f_-(z)]\right.
        + {g_R}^2(T)(1-z)^2\qty[1+\frac{\alpha}{\pi}f_+(z)]
\left. -{g_R}(T)g_L(T)\frac{m_e}{q}z\qty[1+\frac{\alpha}{\pi}f_{+-}(z)]\right\}
\end{equation}
where $m_e$ represents the electron mass, $T=E-m_e$ represents the kinetic energy of recoil electron, $q$ represents incoming neutrino energy, and $z=T/q$ respectively.
Also, $g_L$ and $g_R$ indicate the left-handed and right-handed electron weak couplings, respectively, and $f_{+,-,+-}$ are correction factors from QED\@.
Equation~\eqref{equ:crs_es} corresponds to neutrinos, while the cross section of antineutrinos corresponds to an interchange of $g_L$ and $g_R$.

The charged-current reaction with oxygen (\ce{^{16}O} CC) in the supernova neutrino energy region is a process in which oxygen nuclei interact with neutrinos, leading to a giant resonance in which the entire assembly of nucleons resonates.
An electron or positron is released from the oxygen nucleus for the charged-current reactions, and its nucleus changes to fluorine or nitrogen.
The fluorine and nitrogen have many possible excited states. When the fluorine or nitrogen exceeds the particle emission threshold, particles such as protons, neutrons, and alpha are emitted according to their respective thresholds. What is emitted from the nucleus is determined by which channel the nucleus branches into during deexcitation.
In this simulation, we consider 43 excited states, as listed in Table \ref{tab:nue_excited_state}, with multiple channels considered for a single excited state. The total of 31 channels considered in SKSNSim is detailed in Table \ref{tab:nue_channel}.
The cross section is implemented according 
 to~\cite{PhysRevC.98.034613, SuzukiPrivate}
and provided in SKSNSim\footref{fot:githab}.

In the case of a neutral-current reaction with oxygen (\ce{^{16}O} NC), SK is expected to detect gamma rays emitted during deexcitation, such as from nitrogen or oxygen. In SKSNSim, we only consider states that emit a single gamma-ray during the deexcitation of \ce{^15N} or \ce{^15O}, as \cite{1996PhRvL..76.2629L} specifically consider states of \ce{^15N} or \ce{^15O} generated when a single proton or neutron is emitted from \ce{^16O}.
The cross section is implemented from \cite{1996PhRvL..76.2629L} and \cite{2002PhRvD..66a3007K}.
\begin{deluxetable*}{c|ccccc}
    \centering
    \tablecaption{List of excited states included in SKSNSim.
    \label{tab:nue_excited_state}}
    \tablehead{$^{16}\rm{O}(\nue,e^-\it{N''})\it{N'}$ & \colhead{$0^- (\MeV)$} & \colhead{$1^- (\MeV)$} & \colhead{$2^- (\MeV)$}
     & \colhead{$3^- (\MeV)$} & \colhead{$1^+ (\MeV)$}} 
    \startdata
   & 14.906 & 15.157 & 15.205 & 15.250 & 18.664 \\
   & 27.580 & 20.567 & 19.413 &        & 19.431 \\
   & 27.990 & 23.271 & 21.617 &        & 20.684 \\
   &        & 25.514 & 22.473 &        & 21.937 \\
   &        & 25.718 & 27.255 &        & 22.968 \\
   &        & 26.728 & 27.823 &        & 24.431 \\
   &        & 27.218 & 28.103 &        & 24.858 \\
   &        & 28.141 & 28.922 &        & 25.772 \\
   &        & 28.515 &        &        & 26.593 \\
   &        & 29.200 &        &        & 27.412 \\
   &        & 29.353 &        &        & 27.802 \\
   &        & 29.975 &        &        & 28.082 \\
   &        & 30.250 &        &        & 29.967 \\
   &        & 30.803 &        &        & 31.084 \\
   &        & 31.754 &        &        & 31.693 \\
   &        &        &        &        & 32.981 \\
   \tableline
   $^{16}\rm{O}(\nuebar,e^+\it{N''})\it{N'}$ & \colhead{$0^- (\MeV)$} & \colhead{$1^- (\MeV)$} & \colhead{$2^- (\MeV)$}
     & \colhead{$3^- (\MeV)$} & \colhead{$1^+ (\MeV)$} \\
    \tableline
     & 10.932 & 11.183 & 11.231 & 11.276 & 14.285 \\
     & 23.606 & 16.593 & 15.439 & & 15.052\\
     & 24.016 & 19.297 & 17.643 & & 16.305\\
     & & 21.540 & 18.499 & & 17.558\\
     & & 21.744 & 23.281 & & 18.589\\
     & & 22.754 & 23.849 & & 20.052\\
     & & 23.244 & 24.129 & & 20.479\\
     & & 24.167 & 24.948 & & 21.393\\
     & & 24.541 & & & 22.214\\
     & & 25.226 & & & 23.033\\
     & & 25.379 & & & 23.423\\
     & & 26.001 & & & 23.703\\
     & & 26.276 & & & 25.588\\
     & & 26.829 & & & 26.705\\
     & & 27.780 & & & 27.314\\
     & & & & & 28.602\\
    \enddata
    \tablecomments{This table shows excited states for $^{16}\rm{O}(\nue,e^-\it{N''})\it{N'}$ and $^{16}\rm{O}(\nuebar,e^+\it{N''})\it{N'}$ case. Here, $N'$ is a nucleus in which \ce{^{16}F} or \ce{^{16}N} changed after deexcitaion, and $N''$ is a nucleus emitted by the deexcitation of \ce{^{16}F} or \ce{^{16}N}. $J^{\pi} = 0^-, 1^-, 2^-. 3^-$, and $1^+$ are \ce{^16O} states.}
    \tablerefs{\cite{PhysRevC.98.034613, SuzukiPrivate}}
\end{deluxetable*}
\begin{deluxetable*}{ccccc}
    \tablecaption{List of channels included in SKSNSim.
    \label{tab:nue_channel}}
    \tablehead{ & & \colhead{Reaction: $^{16}\rm{O}(\nue,e^-\it{N''})\it{N'}$} & & } 
     \startdata
    $^{16}\rm{O}(\nue,e^-\gamma)^{16}\rm{F}$ & $^{16}\rm{O}(\nue,e^-n)^{15}\rm{F}$ & & & \\
    $^{16}\rm{O}(\nue,e^-p)^{15}\rm{O}$ & $^{16}\rm{O}(\nue,e^-pn)^{14}\rm{O}$ & &        & \\
    $^{16}\rm{O}(\nue,e^-2p)^{14}\rm{N}$ & $^{16}\rm{O}(\nue,e^-{}^{3}\rm{He})^{13}\rm{N}$ & $^{16}\rm{O}(\nue,e^-\alpha)^{12}\rm{N}$ &        & \\
    $^{16}\rm{O}(\nue,e^-\it{N''})\rm{^{13}C}$ & $^{16}\rm{O}(\nue,e^-\it{N''})\rm{^{12}C}$ & $^{16}\rm{O}(\nue,e^-p\alpha)^{11}\rm{C}$ & $^{16}\rm{O}(\nue,e^-\it{N''})\rm{^{10}C}$ & \\
    $^{16}\rm{O}(\nue,e^-\it{N''})\rm{^{12}B}$ & $^{16}\rm{O}(\nue,e^-\it{N''})\rm{^{11}B}$ & $^{16}\rm{O}(\nue,e^-\it{N''})\rm{^{10}B}$ & $^{16}\rm{O}(\nue,e^-\it{N''})\rm{^{9}B}$ & \\
    $^{16}\rm{O}(\nue,e^-\it{N''})\rm{^{11}Be}$ & $^{16}\rm{O}(\nue,e^-\it{N''})\rm{^{10}Be}$ & $^{16}\rm{O}(\nue,e^-\it{N''})\rm{^{9}Be}$ & $^{16}\rm{O}(\nue,e^-\it{N''})\rm{^{8}Be}$ & $^{16}\rm{O}(\nue,e^-\it{N''})\rm{^{7}Be}$ \\
    $^{16}\rm{O}(\nue,e^-\it{N''})\rm{^{10}Li}$ & $^{16}\rm{O}(\nue,e^-\it{N''})\rm{^{9}Li}$ & $^{16}\rm{O}(\nue,e^-\it{N''})\rm{^{8}Li}$ & $^{16}\rm{O}(\nue,e^-\it{N''})\rm{^{7}Li}$ & $^{16}\rm{O}(\nue,e^-\it{N''})\rm{^{6}Li}$ \\
    $^{16}\rm{O}(\nue,e^-\it{N''})\rm{^{6}He}$ & $^{16}\rm{O}(\nue,e^-\it{N''})\rm{^{5}He}$ & $^{16}\rm{O}(\nue,e^-\it{N''})\rm{^{4}He}$ & $^{16}\rm{O}(\nue,e^-\it{N''})\rm{^{3}He}$ & \\
    $^{16}\rm{O}(\nue,e^-\it{N''})\rm{^{3}H}$ & $^{16}\rm{O}(\nue,e^-\it{N''})\rm{^{2}H}$ & & & \\
    \tableline
     & & \colhead{Reaction: $^{16}\rm{O}(\nuebar,e^+\it{N''})\it{N'}$} & & \\
    \tableline
    $^{16}\rm{O}(\nuebar,e^+\gamma)^{16}\rm{N}$ & $^{16}\rm{O}(\nuebar,e^+n)^{15}\rm{N}$ & $^{16}\rm{O}(\nuebar,e^+2n)^{14}\rm{N}$ & $^{16}\rm{O}(\nuebar,e^+\it{N''})\rm{^{13}N}$ & \\
    $^{16}\rm{O}(\nuebar,e^+pn)\rm{^{14}C}$ & $^{16}\rm{O}(\nuebar,e^+{}^{3}\rm{H})\rm{^{13}C}$ & $^{16}\rm{O}(\nuebar,e^+\it{N''})\rm{^{12}C}$ & $^{16}\rm{O}(\nuebar,e^+\it{N''})\rm{^{11}C}$ & $^{16}\rm{O}(\nuebar,e^+\it{N''})\rm{^{10}C}$ \\
    $^{16}\rm{O}(\nuebar,e^+\it{N''})\rm{^{14}B}$ & $^{16}\rm{O}(\nuebar,e^+\it{N''})\rm{^{13}B}$ & $^{16}\rm{O}(\nuebar,e^+\alpha)\rm{^{12}B}$ & $^{16}\rm{O}(\nuebar,e^+\it{N''})\rm{^{11}B}$ & $^{16}\rm{O}(\nuebar,e^+\it{N''})\rm{^{10}B}$ \\ $^{16}\rm{O}(\nuebar,e^+\it{N''})\rm{^{9}B}$ & $^{16}\rm{O}(\nuebar,e^+\it{N''})\rm{^{8}B}$ & & & \\
    $^{16}\rm{O}(\nuebar,e^+\it{N''})\rm{^{12}Be}$ & $^{16}\rm{O}(\nuebar,e^+\it{N''})\rm{^{11}Be}$ & $^{16}\rm{O}(\nuebar,e^+\it{N''})\rm{^{10}Be}$ & $^{16}\rm{O}(\nuebar,e^+\it{N''})\rm{^{9}Be}$ & $^{16}\rm{O}(\nuebar,e^+\it{N''})\rm{^{8}Be}$ \\ $^{16}\rm{O}(\nuebar,e^+\it{N''})\rm{^{7}Be}$ & & & & \\
    $^{16}\rm{O}(\nuebar,e^+\it{N''})\rm{^{9}Li}$ & $^{16}\rm{O}(\nuebar,e^+\it{N''})\rm{^{8}Li}$ & $^{16}\rm{O}(\nuebar,e^+\it{N''})\rm{^{7}Li}$ & $^{16}\rm{O}(\nuebar,e^+\it{N''})\rm{^{6}Li}$ & \\
    $^{16}\rm{O}(\nuebar,e^+\it{N''})\rm{^{6}He}$ & $^{16}\rm{O}(\nuebar,e^+\it{N''})\rm{^{4}He}$ & $^{16}\rm{O}(\nuebar,e^+\it{N''})\rm{^{3}He}$ & & \\
    $^{16}\rm{O}(\nuebar,e^+\it{N''})\rm{^{3}H}$ & $^{16}\rm{O}(\nuebar,e^+\it{N''})\rm{^{2}H}$ & & & \\
    \enddata
    \tablecomments{This table shows channels for $^{16}\rm{O}(\nue,e^-\it{N''})\it{N'}$ and $^{16}\rm{O}(\nuebar,e^+\it{N''})\it{N'}$ case. Here, $N'$ is a nucleus in which \ce{^{16}F} or \ce{^{16}N} changed after deexcitation, and $N''$ is a nucleus emitted by deexcitation of \ce{^{16}F} or \ce{^{16}N}.}
    \tablerefs{\cite{PhysRevC.98.034613, SuzukiPrivate}.}
\end{deluxetable*}

\subsection{Output Format}\label{subsec:output}

This simulation results contain details regarding the interaction channel and particle kinematics for every event. Each neutrino event includes information about the particle type, time, and position of the neutrino interaction, as well as the particle type, direction, and energy of generated particles.
This information is available in two data formats: the SK custom format employed in the SK offline analysis package and NUANCE format\footnote{The documentation is available on \url{http://neutrino.phy.duke.edu/nuance-format/}.}.
For SK, the output information from SKSNSim is transferred to the detector simulation, including simulation of the trigger system, for further assessment of the detector response through the analysis pipeline. Following the simulation, the expected number of events for each interaction is recorded. Specific examples will be provided in the following section.

\section{Demonstration}\label{sec:demo}

In this section, we describe demonstrations of SKSNSim.
It generates observable particles in water Cherenkov detectors and can pass information about these particles to detector simulators in two modes: the SN burst and the DSNB mode.
Section~\ref{subsec:generated_events} provides examples of events generated by neutrino interactions in the SN burst mode and positron events generated by IBD in the DSNB mode.
Section~\ref{subsec:use_sk} explains how the generated events can be used for SK analysis.\@

\subsection{Generated Neutrino Events}\label{subsec:generated_events}

This section shows the simulation result of an SN burst and a DSNB using SKSNSim.
The process begins by calculating the number of events for each interaction, followed by defining the kinematics for each event. Users have the flexibility to change the direction of the neutrino in each event and distance from an supernova to a detector.
In both modes, the neutrino energy is determined by a random number that follows the energy distribution defined by the model, and the interaction vertex is generated uniformly inside the detector. Furthermore, using the previously determined energy and direction of the neutrino, the kinematics of outgoing particles, such as positrons in the IBD interaction, are determined based on their respective differential cross section.

\begin{figure*}
    \gridline{\fig{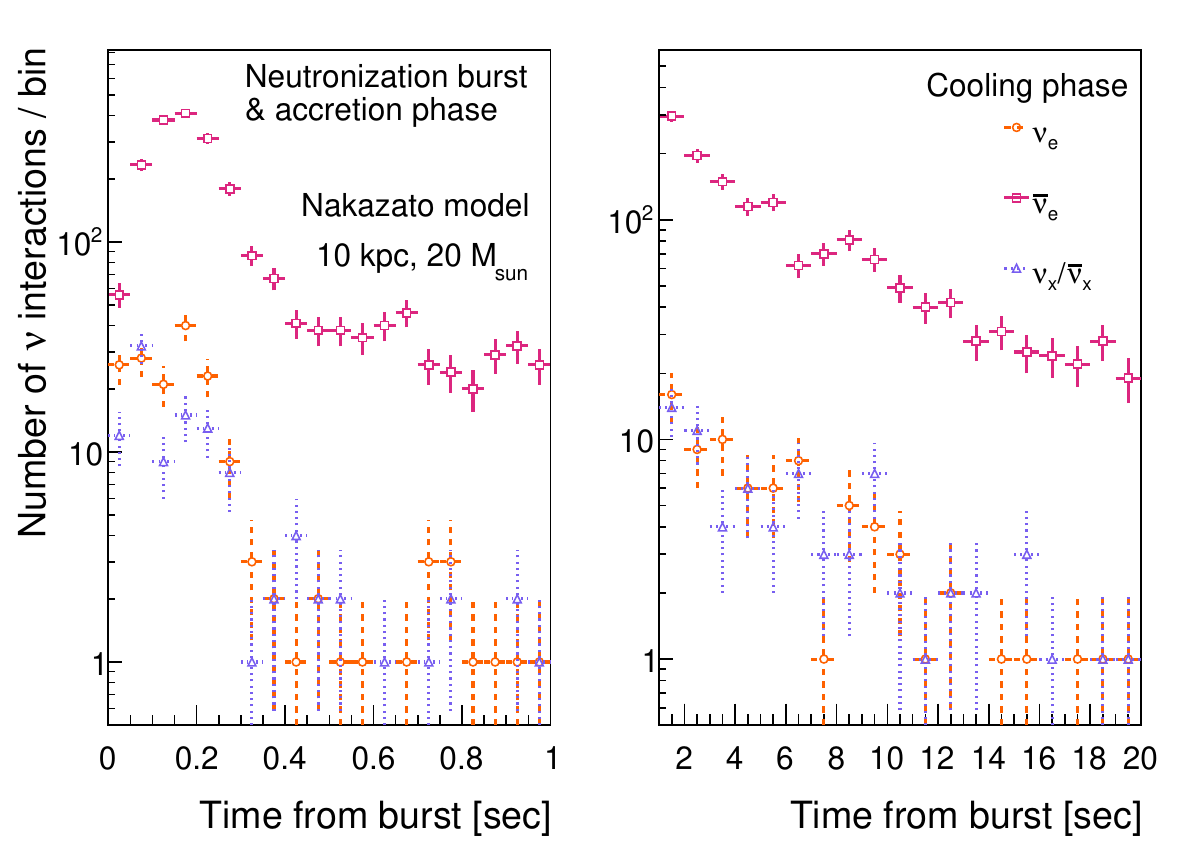}{0.33\textwidth}{(a)}
          \fig{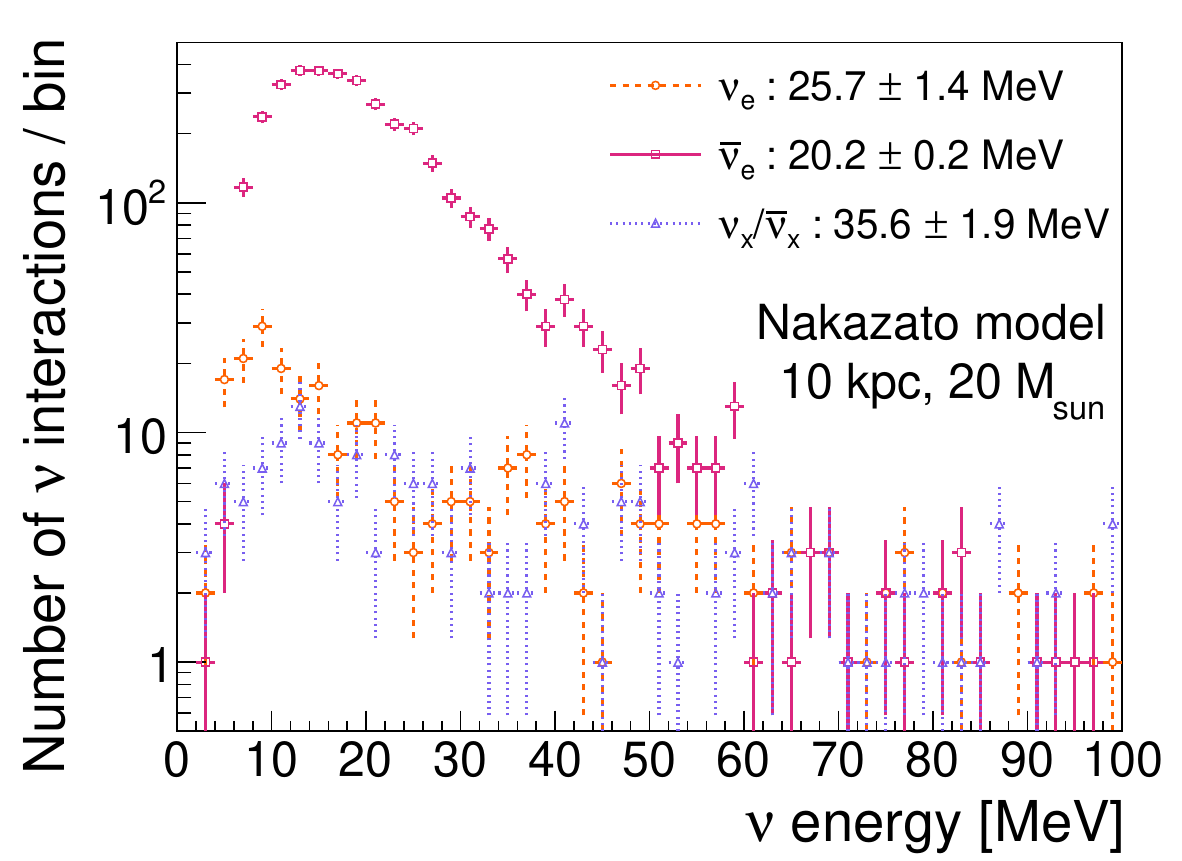}{0.33\textwidth}{(b)}
          \fig{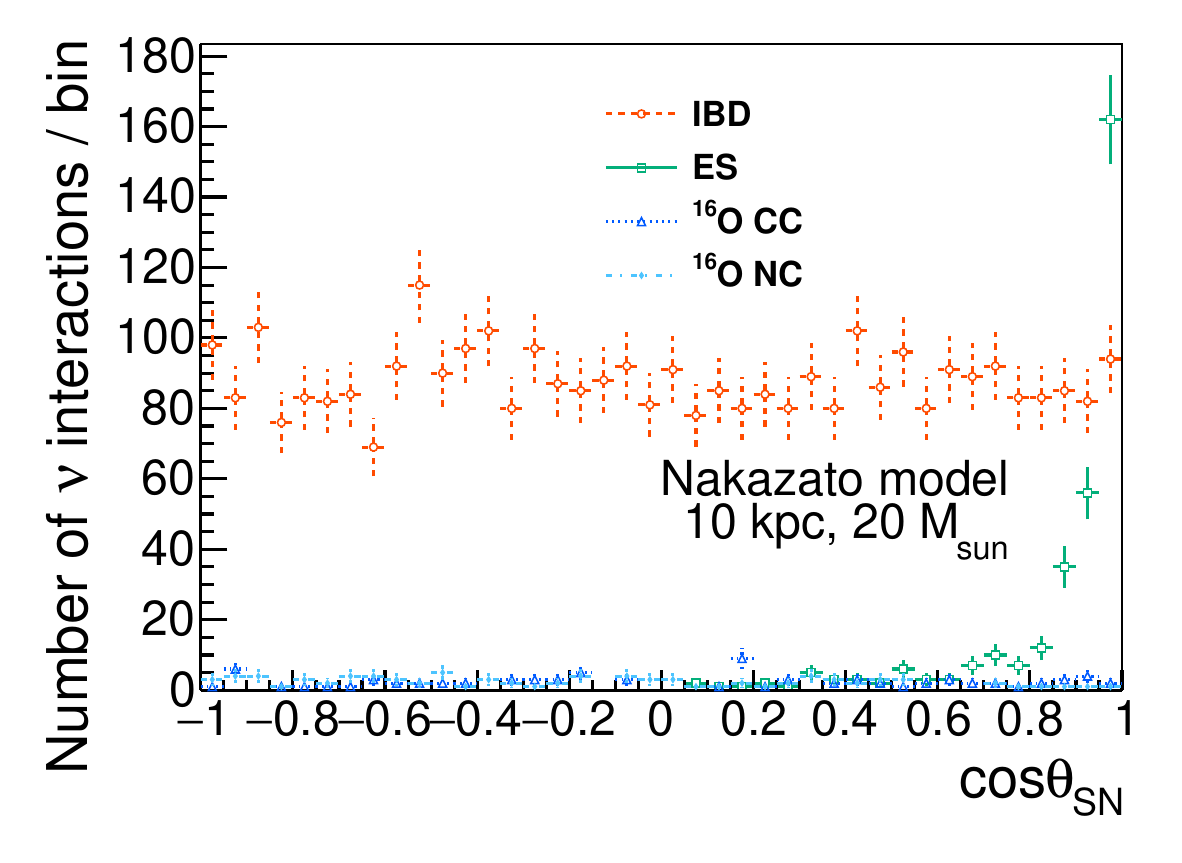}{0.33\textwidth}{(c)}
          }
          \caption{Distributions of neutrino events generated by a typical SN burst. These distributions show true kinematics without detector response. (a) Timing distributions of generated neutrino events. The left focuses on the neutronization burst and accretion phases. The right figure shows that of cooling phase. (b) Energy distribution of generated neutrino events integrated over all phases. The numbers in the legend are the mean energy for each neutrino type. (c) $\theta_{\rm{SN}}$ distribution between neutrinos and positrons in the IBD distribution, between neutrinos and electrons in the ES distribution, between neutrinos and electrons or positrons in the \ce{^{16}O} CC, and between neutrinos and gamma-rays in the \ce{^{16}O} NC. The model is 
    \cite{2013ApJS..205....2N}, for which the parameters of the progenitor are mass of $M = 20~\Msun$, shock revival time of $200~\rm{ms}$, metallicity of $Z=0.02$, and distance of $d = 10~\kpc$. We assume no neutrino oscillation in these plots.\label{fig:demo_snburst_nakazato}}
\end{figure*}

Figure~\ref{fig:demo_snburst_nakazato} shows the time, energy, and angular distributions of neutrino events interacting with water in the case of an SN burst. In the SN burst mode, the type of neutrino interaction, its time and vertex, as well as the type of generated particle, its direction, and energy, are stored as event information for a single supernova explosion. $\theta_{\rm{SN}}$ represents the angle between neutrinos and a generated particle for each interaction.
This case assumes the Nakazato model~\citep{2013ApJS..205....2N}.

\begin{figure}
    \centering
    \includegraphics[width=0.4\textwidth]{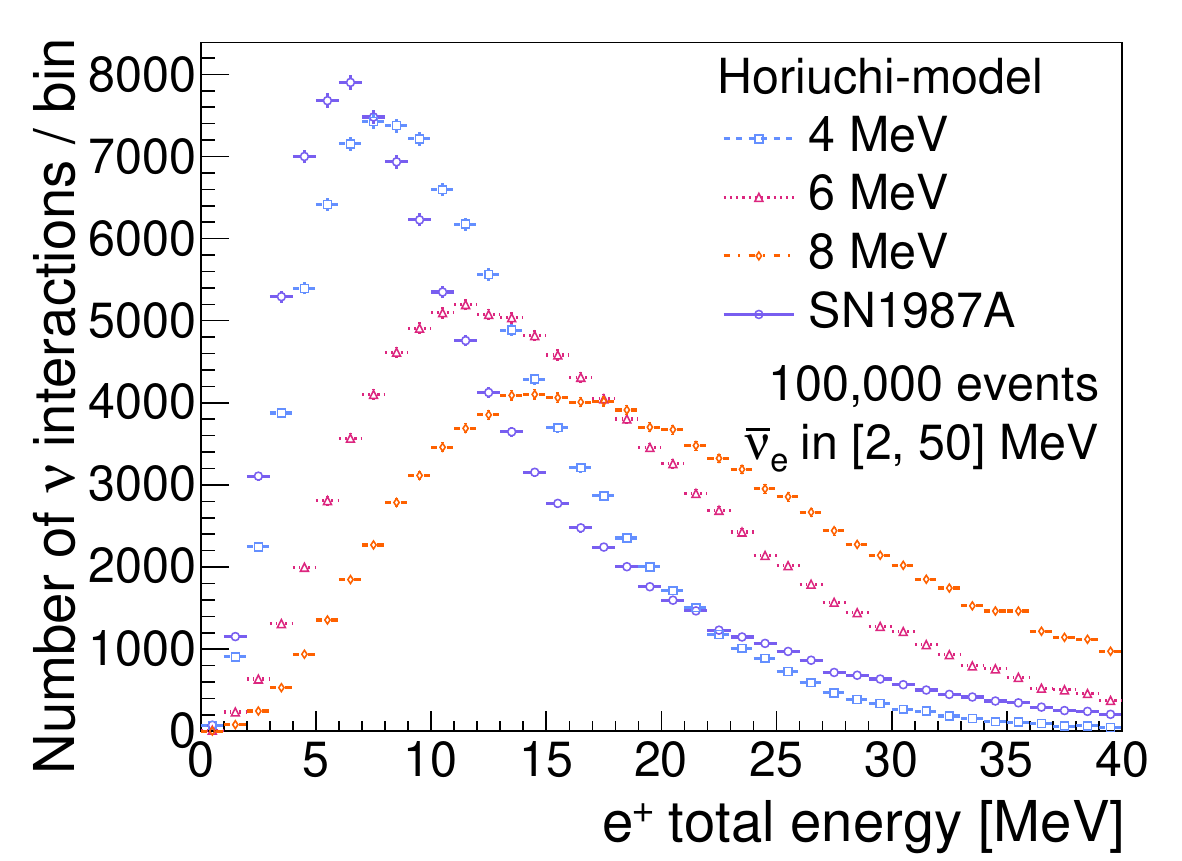}
    \caption{Energy spectra of positrons generated by DSNB with water via the IBD channel. Here, the calculation of \cite{2009PhRvD..79h3013H} is used for the DSNB model, and \cite{2003PhLB..564...42S} for the IBD cross section.\label{fig:dsnb_generated_spectrum}}
\end{figure}

Figure~\ref{fig:dsnb_generated_spectrum} shows an example of positron energy distribution from 100,000 IBD events generated by DSNB under several assumptions in \cite{2009PhRvD..79h3013H}.
The direction of neutrinos is determined isotropically.

\subsection{Usage for Super-Kamiokande}\label{subsec:use_sk}

This section briefly explains the usage in the SK analysis.
The simulator was originally designed for the analysis in SK, especially for sensitivity calculations to the SN burst~(Y.Kashiwagi et al. 2024, in preparation) 
and the optimization of the DSNB analysis~\citep{2023ApJ...951L..27H}.
The pipeline for the analysis consists of a neutrino interaction part (SKSNSim), the full detector simulation of generated particles, and the detailed analysis for specific studies.
The SK software package holds an internal data format that is consistent throughout the tools processing pipeline using the ROOT package~\citep{1997NIMPA.389...81B}.
The SK detector simulator simulates the detector response for particles generated in neutrino interactions, which are the output of SKSNSim.

In the case of an SN burst, the simulation flow is straightforward.
SKSNSim generates the neutrino event information, and the detector simulator processes the output for each neutrino event.
After that, the time series of neutrino events is considered.
SK observes Cherenkov photons emitted by charged particles generated in a neutrino interaction.
These photons may overlap with each other in neutrino events, especially under high event rates during the early phase.
For this reason, the detected photons are ordered by arrival time independently of the neutrino event, and after being combined with the continuous detector noise, a trigger simulation is applied.
After being reconstructed as SK events by trigger simulation, the timing profile and energy of neutrino emission from the SN burst, and the direction of the original supernova source are estimated~(Y.Kashiwagi et al. 2024, in preparation).

In contrast, the DSNB process uses a different procedure~\citep{2023ApJ...951L..27H}.
In the step of SKSNSim, we first generate the necessary number of positron events, assuming that the positron energy distribution is flat, and then determine the original $\nuebar$ kinematics by throwing a random number according to the IBD differential cross section for each positron event.
Each positron event is weighted according to the original neutrino energy distribution when making the positron energy distribution in the analysis pipeline.
This method is called the ``re-weighting'' method. It can be applied with any $\nuebar$ spectrum, for example, $\nuebar$ from reactors, because only the IBD interaction of $\nuebar$ is considered.

\section{Summary}\label{sec:summary}

SKSNSim is software to simulate neutrino interaction events of supernova bursts and diffuse supernova neutrino background.
We have developed SKSNSim originally for SK and modified it for general purposes and public use.
In this paper, we described the basic structures.
It outputs the interaction position and the outgoing particles' type, direction, and energy according to the differential cross section of neutrino interactions in the water.
Based on this output information, the detector simulator traces the Cherenkov photons emitted from the interacted particles.
We demonstrated SKSNSim in both SN burst and DSNB and explained how it is used in these analyses.
It applies to any water-based detector, for example, the Hyper-Kamiokande detector, which is now under construction~\citep{Neutrino2022HyperK}.

\section*{Acknowledgement}\label{sec:acknowledgement}

We would like to thank Y.~Kashiwagi for providing the text data of the SN burst models, and T.~Suzuki for many useful suggestions about oxygen-excited states after neutrino interactions.
We are also grateful for the helpful discussions with K. Scholberg, M. Nakahata, H. Sekiya, M. Ikeda, and G. Pronost.
This work is supported by JSPS KAKENHI Grant Number JP23KJ1609, JP23KJ0890, JP20J20189, and JP20H00162. 
This work was supported by JST, the establishment of university fellowships towards the creation of science technology innovation, Grant Number JPMJFS2112. 

\bibliography{refsksnsim}{}
\bibliographystyle{aasjournal}


\listofchanges


\end{document}